\documentclass[12pt,preprint]{aastex}
\usepackage{longtable}
\usepackage{pdflscape}	

\begin{document}

\title{DE CVn: an eclipsing post-common envelope binary with a circumbinary disk and a giant planet}

\author{Han Z.-T\altaffilmark{1,2,3,4}, Qian S.-B\altaffilmark{1,2,3,4,5}, Zhu L.-Y\altaffilmark{1,3,4,5}, Zhi Q.-J\altaffilmark{2}, Dong A.-J\altaffilmark{2}, Soonthornthum B.\altaffilmark{6}, Poshyachinda S.\altaffilmark{6}, Sarotsakulchai T.\altaffilmark{1,4,6}, Fang X.-H\altaffilmark{1,3,5}, Wang Q.-S\altaffilmark{1,3,5}, Irina Voloshina\altaffilmark{7}}

\singlespace

\altaffiltext{1}{Yunnan Observatories, Chinese Academy of Sciences (CAS), P. O. Box 110, 650216 Kunming, China; zhongtaohan@ynao.ac.cn}
\altaffiltext{2}{Guizhou Provincial Key Laboratory of Radio Data Processing, School of Physics and Electronic Sciences, Guizhou Normal University, Guiyang 550001, China}
\altaffiltext{3}{Key Laboratory of the Structure and Evolution of Celestial Objects, Chinese Academy of Sciences, P. O. Box 110, 650216 Kunming, China}
\altaffiltext{4}{Center for Astronomical Mega-Science, Chinese Academy of Sciences, 20A Datun Road, Chaoyang District, Beijing, 100012, P. R. China}
\altaffiltext{5}{University of Chinese Academy of Sciences, Yuquan Road 19\#, Sijingshang Block, 100049 Beijing, China}
\altaffiltext{6}{National Astronomical Research Institute of Thailand, Ministry of Science and Technology, Bangkok, Thailand}
\altaffiltext{7}{Sternberg Astronomical Institute, Moscow State University, Universitetskij prospect 13, Moscow 119992, Russia}

\begin{abstract}
We present a timing analysis of the eclipsing post-common envelope binary (PCEB) DE CVn.
Based on new CCD photometric observations and the published data, we found that the orbital period in DE CVn has a cyclic period oscillation with an amplitude of $28.08$ s and a period of $11.22$ years plus a rapid period decrease at a rate of $\dot{P}=-3.35\times10^{-11}ss^{-1}$. According to the evolutionary theory,
secular period decreases in PCEBs arise from angular momentum losses (AMLs) driven by gravitational radiation (GR) and magnetic braking (MB).
However, the observed orbital decay is too fast to be produced by AMLs via GR and MB, indicating that there could be other AML mechanism.
We suggest that a circumbinary disk around DE CVn may be responsible for the additional AML. The disk mass was derived as a few$\times$$10^{-4}$-$10^{-3}$$M_{\odot}$
, which is in agreement with that inferred from previous studies in the order of magnitude.
The cyclic change is most likely result of the gravitational perturbation by a circumbinary object due to the Applegate's mechanism
fails to explain such a large period oscillation. The mass of the potential third body is calculated as $M_{3}\sin{i'}=0.011(\pm0.003)M_{\odot}$.
Supposing the circumbinary companion and the eclipsing binary is coplanar, its mass would correspond to a giant planet. This hypothetical giant planet is moving in a circular orbit of radius $\sim5.75(\pm2.02)$ AU around its host star.

\end{abstract}


\section{Introduction}
Post-common envelope binaries (PCEBs) consisting of a white dwarf or hot subdwarf B/O (sdB) primary and a low-mass stellar or brown dwarf secondary are an important class of highly evolved binaries. These binary systems are survivors of a common envelope (CE) phase (e.g., Paczynski 1976; Webbink 2008, Zorotovic et al. 2010), and their orbital periods are in a range of a few hours to days (Rebassa-Mansergas et al. 2012).
Subsequently, PCEBs continue to shrink their orbits due to angular momentum loss (AML) from the system, eventually forming semi-detached cataclysmic variable stars (CVs). Eclipsing PCEBs allows for accurate timing measurements because the brightness and radius of both white dwarf and red dwarf are very different (Parsons et al. 2010).
This provide a good opportunity to detect the evolution of systems and circumbinary planets.
Previous studies have suggested that almost all eclipsing PCEBs with a baseline of more than 5 years display obvious period changes (see Zorotovic \& Schreiber 2013, for more details of eclipsing examples).
In general, the secular period decrease arises from the AML via gravitational radiation (GR) (Paczy{\'n}ski 1967) and magnetic braking (MB) (Verbunt \& Zwaan 1981), but the periodic period variation can be interpreted by the Applegate mechanism (Applegate 1992) or the light-travel-time (LTT) effect from the influence of an unseen substellar object (e.g., Guinan \& Ribas 2001; Lee et al. 2009; Qian et al. 2009a,b, 2010a, 2011; Beuermann et al. 2010, 2011; Potter et al. 2011; Marsh et al. 2014). Therefore, by the precise eclipse timings, the evolutionary state of these binaries (e.g., PCEBs and CVs) can be ascertained and the theories of AML can be tested. Also, they are very ideal targets that can be used to search for circumbinary planets (e.g., Qian et al. 2015, 2016). The planets orbiting PCEBs are particularly important and interesting because the host stars have evolved past the CE phase.
Recently, by using the timing method, a lot of exoplanets orbiting PCEBs have been detected, such as DP Leo (Beuermann et al. 2011), HU Aqr (Qian et al. 2011; Go\'{z}dziewski et al. 2015), NN Ser (Marsh et al. 2014), QS Vir (Qian et al. 2010b; Almeida \& Jablonski 2011), NY Vir (Qian et al. 2012b; Lee et al. 2014), RR Cae (Qian et al. 2012a), HS0705+6700 (Qian et al. 2013), and DV UMa (Han et al. 2017a).
Some of the claimed circumbinary planetary systems have been tested by detailed dynamical stability analyses, and most such systems have been found to be unfeasible (e.g. Hinse et al. 2012; Horner et al. 2012; Wittenmyer et al. 2013). However, Marsh et al. (2014) pointed out that such works regarding the stability are flawed and require revision. Therefore, an updated method is needed to assess the long-term dynamical stability of the proposed circumbinary planet systems.

Using the timing method our group since 2009 tried to search for extrasolar planets around the white dwarf binaries and to detect the secular evolution of these systems (e.g., Dai et al. 2009, 2010; Qian et al. 2009, 2010, 2011, 2012, 2015, 2016; Han et al. 2015, 2016, 2017a,b,c,d). Here, we report the photometric observations of the eclipsing PCEB DE CVn since 2009 and obtain new mid-eclipse times. This object in the ROSAT catalogue was first considered to be an X-ray source by Voges et al. (1999). A study by Robb \& Greimel (1997) also discovered that it is an eclipsing white-dwarf binary. Based on the light curve and its photometric property, they measured the orbital period and the eclipse depth.
Further observations by Holmes \& Samus (2001) concluded that the eclipse depths depend on the colours. Later, van den Besselaar et al. (2007) presented the photometric and spectroscopic observations of DE CVn that given an accurate ephemeris and derived the system parameters. The orbital ephemeris was improved by several previous authors based on new eclipse times of DE CVn, but no sign of any period variations were claimed (Parsons et al. 2010; Lohr et al. 2014). The main reasons are that most of the published timings have larger errors and the observational baseline is still quite short. In this paper, further accurate eclipse timings were presented and the detailed analysis of period changes in DE CVn was made. We find a secular decrease together with a periodic change in the orbital period. At the end, we discuss the AML mechanisms and the presence of a giant planet.

\section{Observations and Data Reduction}

To obtain additional eclipse timings, new photometry of DE CVn were taken using some telescopes and instruments from March 2009 to May 2017.
They were: the 2.4 m telescope mounted both a VersArray 1300B CCD camera in 2009 and YFOSC (Yunnan Faint Object Spectrograph and Camera, 2K$\times$4K) after 2012 at the Lijiang observational station of Yunnan Observatories (YNOs); the 60 cm and the 1.0 m reflecting telescopes attached Andor DW436 2K CCD cameras at YNOs; the 85 cm and
the 2.16 m telescopes at Xinglong Station administered by National Astronomical Observatories, Chinese Academy of Sciences (NAO), the detectors are the Andor DW436 1K CCD
camera and the PI 1274$\times$1152 TE CCD, respectively; and the 2.4 m Thai National Telescope (TNT) of National Astronomical Research Institute Of Thailand (NARIT) equipped with an ULTRASPEC fast camera.

First of all we made bias, dark and flat-field corrections to the raw images. Then we use the IRAF software with aperture photometry to reduce these photometric data.
The comparison star and check star of no intrinsic variability were chosen to perform the differential photometry.
The comparison star is 2MASS J13265966+4533035 ($13^{h}26^{m}59.68^{s}$, $+45^{\circ}33^{\prime}03.53^{\prime\prime}$, J2000.0), and the check star is GSC 03460-00601 ($13^{h}27^{m}04.37^{s}$, $+45^{\circ}35^{\prime}30.44^{\prime\prime}$, J2000.0). A summary of the observation instruments is listed in Table 1. Four primary eclipses obtained with 2.4 m and 85 cm are shown in Figure 1. Clearly, the eclipses of white dwarf are both steep and distinct so that they can be timed precisely ($<$ 10 s).
The mid-eclipse times are determined by averaging four times, which are the start and end times of white dwarf's ingress and egress, respectively. To measure the times of these contact points, we used three straight lines to fit the section around ingress (or egress). The out-of-eclipse data were fitted by the first line, the flat bottom of eclipse light curves was fitted by the second line, and the steep slope in the eclipse profile corresponding to the ingress (or egress) data was fitted by the third line. The points of intersection of all these lines were regarded as the four times needed. The exposure time for the different nights was adopted as 5 s, 6 s and 15 s, respectively.
We obtained twenty-two mid$-$eclipse times by fitting new data. Their errors were defined as the standard deviation values during calculation, depending on the signal-to-noise ratios and integration times at the time of the observation. Apart from these observations, we also found that the AAVSO (American Association of Variable Star Observers) data contain many eclipsing light curves. Note that most of observations were obtained between March and May 2010.
Such data were first prepared by extracting the eclipsing profiles before the measurements. Applying these eclipse data, then, seven accurate eclipse timings were determined by using same method above. Here we only used a subset of these data to produce the mid-eclipse times, the reason is that the uncertainties of another data are too large. These available timings and the corresponding information were collected in Table 1.

\begin{landscape}
\begin{table*}
\caption{New mid-eclipse times of DE CVn.}
 \small
 \begin{tabular}{llllllllll}\hline\hline
Date           &Min.(HJD)             &Min.(BJD)               &E        &O-C              & Err        &$T_{exp}(s)$   &$N_{obs}$          &Telescopes       &Fil.    \\\hline
2009 Mar 24    &2454915.13219         &2454915.13297          &5851	    &-0.00021          & 0.00005    &5.0           &590            &2.4m             &N             \\
2009 Mar 24    &2454915.13221         &2454915.13299          &5851	    &-0.00019          & 0.00005    &5.0           &111            &1m               &R             \\
2009 Apr 21    &2454943.17087         &2454943.17165          &5928      &-0.00026         & 0.00005    &5.0           &468            &2.4m             &V             \\
2009 May 02    &2454954.09513         &2454954.09591          &5958	    &-0.00018          & 0.00010    &6.0           &334            &85cm             &V             \\
2010 Mar 22    &2455277.45103         &2455277.45181          &6846	    &0.00001 	         & 0.00010    &-           &429            &AAVSO            &V             \\
2010 Mar 23    &2455278.54331         &2455278.54409          &6849	    &-0.00012	         & 0.00010    &-           &1276           &AAVSO            &V              \\
2010 Mar 23    &2455278.54338         &2455278.54416          &6849	    &-0.00005	         & 0.00010    &-           &745            &AAVSO            &V             \\
2010 Mar 30    &2455285.82611         &2455285.82689          &6869	    &-0.00011	         & 0.00010    &-           &443            &AAVSO            &V             \\
2010 Apr 01    &2455287.64688         &2455287.64766          &6874	    &-0.00004	         & 0.00010    &-           &2321           &AAVSO            &V              \\
2010 Apr 23    &2455309.49525         &2455309.49603          &6934	    &-0.00003	         & 0.00010    &-           &2331           &AAVSO            &V              \\
2010 May 04    &2455320.41949         &2455320.42027          &6964	    &0.00004 	         & 0.00010    &-           &1002           &AAVSO            &V              \\
2011 Jan 02    &2455564.39272         &2455564.39350          &7634	    &-0.00008          & 0.00010    &6.0           &315            &1m               &R             \\
2011 Mar 20    &2455641.22627         &2455641.22705          &7845	    &0.00008           & 0.00015    &15.0           &442            &60cm             &N             \\
2012 Feb 14    &2455972.22889         &2455972.22967          &8754	    &0.00006           & 0.00005    &5.0           &207            &2.4m             &N              \\
2012 Feb 15    &2455973.32124         &2455973.32202          &8757	    &-0.00001          & 0.00005    &5.0           &165            &2.4m             &N              \\
2012 Apr 05    &2456023.20833         &2456023.20911          &8894	    &-0.00001          & 0.00010    &6.0           &217            &1m               &N              \\
2012 Apr 15    &2456033.04018         &2456033.04096          &8921	    &0.00008           & 0.00010    &6.0           &173            &60cm             &N              \\
2012 Apr 20    &2456038.13812         &2456038.13890          &8935	    &0.00007           & 0.00015    &15.0           &181            &60cm             &N              \\
2012 Apr 23    &2456041.05110         &2456041.05188          &8943	    &-0.00007          & 0.00015    &15.0           &416            &60cm             &N              \\
2012 Jun 06    &2456085.11201         &2456085.11279          &9064	    &-0.00001          & 0.00005    &5.0           &212            &2.4m             &N              \\
2012 Dec 31    &2456293.39964         &2456293.40042          &9636	    &-0.00007          & 0.00005    &5.0           &510            &2.4m             &N              \\
2013 Mar 04    &2456356.39608         &2456356.39686          &9809	    &0.00027           & 0.00010    &6.0           &121            &2.16m            &R              \\
2013 Mar 27    &2456379.33653         &2456379.33731          &9872	    &-0.00006          & 0.00010    &6.0           &202            &1m               &N              \\
2015 Jan 30    &2457053.35798         &2457053.35875          &11723	  &-0.00049          & 0.00010    &6.0           &404            &Thai2.4m         &N               \\
2015 Mar 19    &2457101.06008         &2457101.06085          &11854	  &-0.00065          & 0.00010    &6.0           &447            &85cm             &N               \\
2015 Apr 01    &2457114.16896         &2457114.16973          &11890	  &-0.00078          & 0.00010    &6.0           &282            &1m               &N               \\
2016 Jun 07    &2457547.13016         &2457547.13094          &13079	  &-0.00122          & 0.00010    &6.0           &453            &85cm             &N               \\
2017 May 08    &2457882.13782         &2457882.13861          &13999	  &-0.00171          & 0.00010    &6.0           &112            &1m               &B               \\
2017 May 08    &2457882.13767         &2457882.13846          &13999	  &-0.00186          & 0.00010    &6.0           &113            &1m               &V               \\
\hline
\end{tabular}
\end{table*}
\end{landscape}

\begin{figure}[!h]
\begin{center}
\includegraphics[width=16cm]{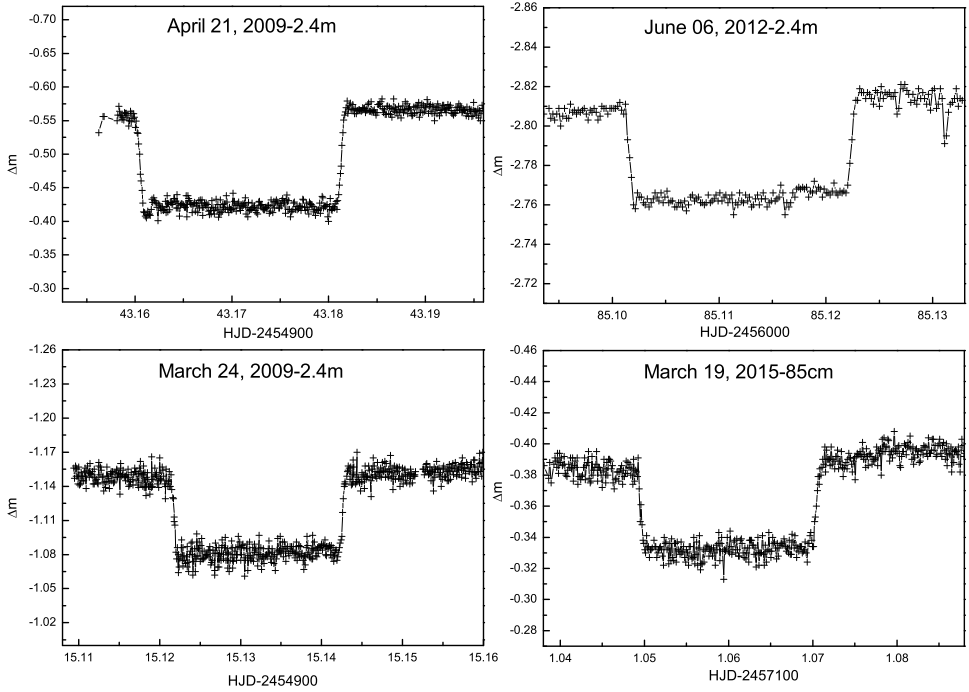}
\caption{Four selected primary eclipses of DE CVn observed with the 85cm and 2.4m telescopes in China.}
\end{center}
\end{figure}

\begin{figure}[!h]
\begin{center}
\includegraphics[width=16cm]{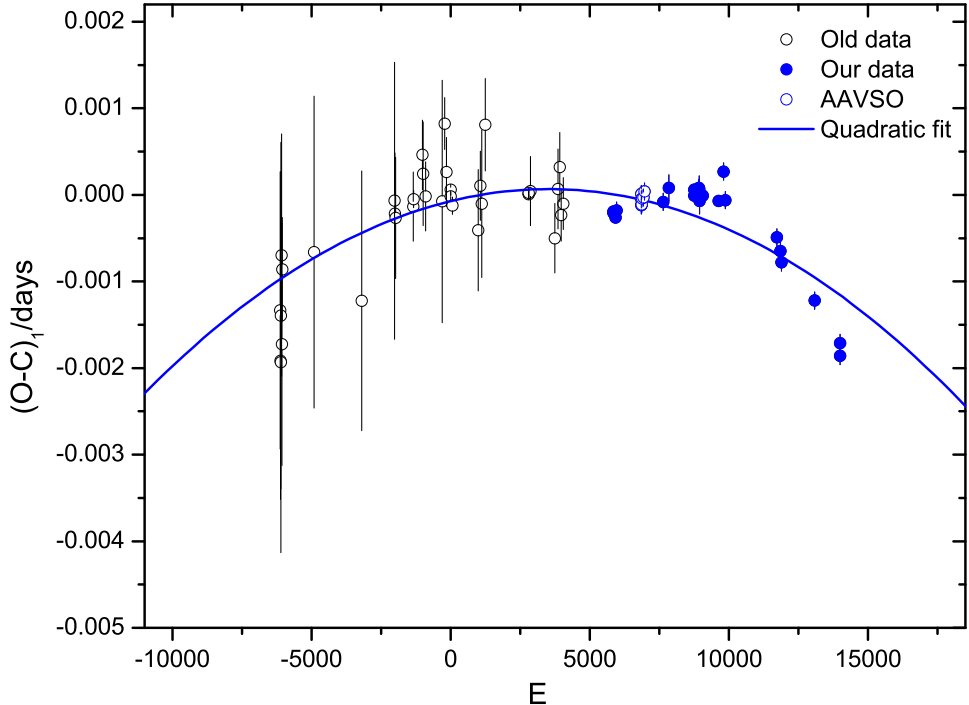}
\caption{Updated O-C diagram of DE CVn using new and historical mid-eclipse times.  The black open circles denotes all published timings, while blue circles denote new eclipse timings (open and solid). The blue solid line represents the best-fit quadratic ephemeris. }
\end{center}
\end{figure}

\begin{figure}[!h]
\begin{center}
\includegraphics[width=16cm]{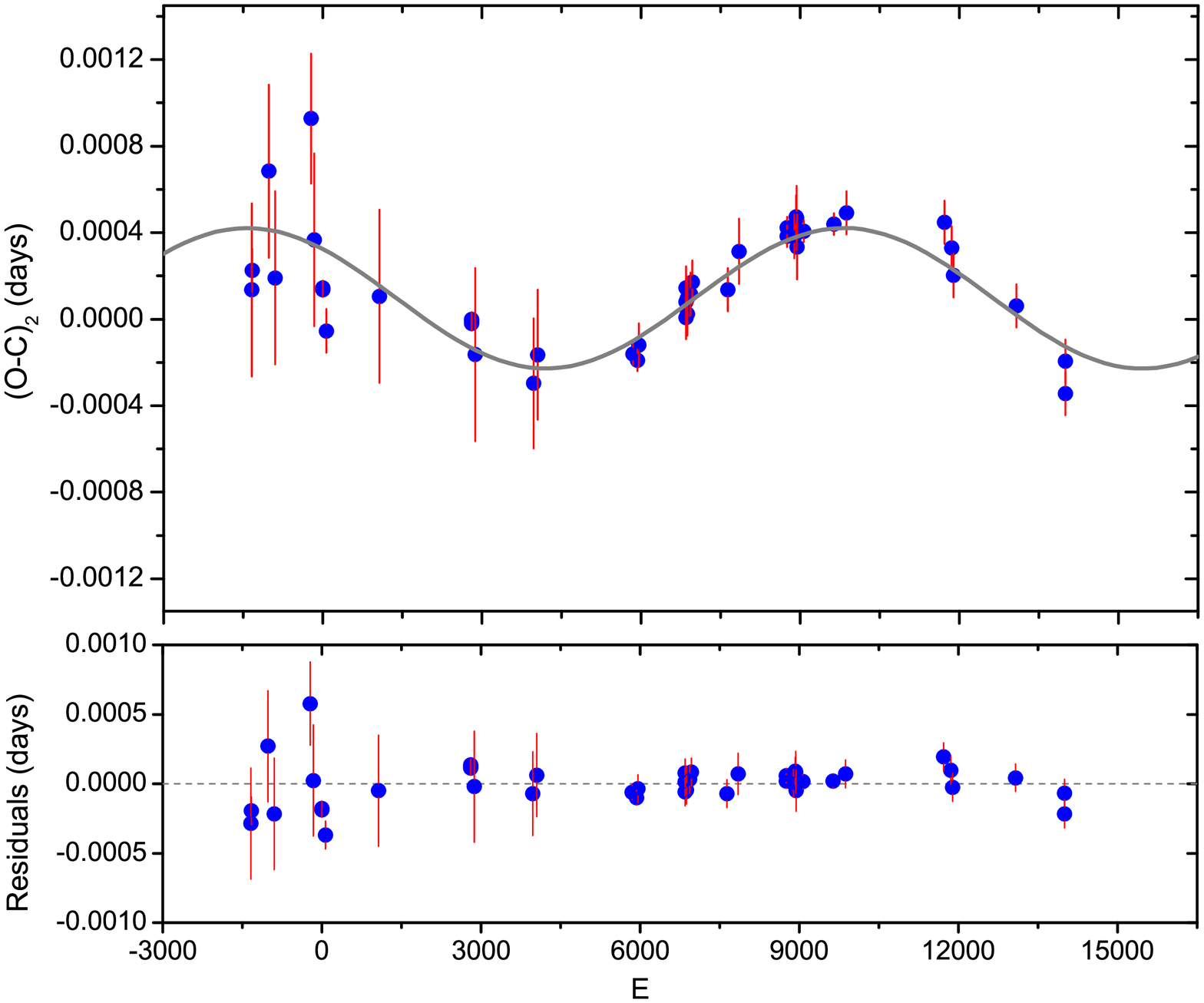}
\caption{$(O-C)_2$ plot extracted from Figure 2 with respect to the sinusoidal curve is displayed in the top panel where a cyclic period wiggle is clearly visible. After the periodic change was removed, the residuals are shown in the low panel.}
\end{center}
\end{figure}

\section{Analysis and Results}

The eclipse timings of DE CVn have been presented previously and the period changes also have been studied (e.g., Robb \& Greimel 1997; Tas et al. 2004; Van den Besselaar et al. 2007; Parsons et al. 2010; Lohr et al. 2014). Since there are only a few precise timings in the historical data, however, they had not found any period changes.
New observations shown in Table 1, coupled with historical data from the literatures, present a newest $O-C$ diagram (see Figure 2). All $O-C$ values were determined using the orbital ephemeris of Parsons et al. (2010),
\begin{equation}
Min.I(BJD)=2452784.554043(1)+0.3641393156(5)\times E,
\end{equation}
where E is the cycle number. The time span of the latest $O-C$ curve has increased to $\sim$ 20 years.

We first apply a linear ephemeris to represent the $O-C$ plot. However, all observed timings show significant deviations from this ephemeris, and the fitting residuals reveal a long-term period decrease. Using a quadratic ephemeris to the data leads to a much better fit and is displayed in Figure 2.
Such fit described the general trend of the $O-C$ curve can be written as
\begin{equation}
(O-C)_1=\Delta{T_0}+\Delta{P_0}\times{E}+{\beta}E^{2}.
\end{equation}
The explanations and the derived values of the fitting parameters are listed in Table 2.
In the process of analysis, we used the Levenberg-Marquart (LM) method to fit the $O-C$ plot. 
The weight for each data point is inversely proportional to the size of its error. The errors of fitting parameter are the formal errors derived by LM technique. To describe the goodness-of-fit of each model, the chi-squared values ($\chi^{2}$) are calculated to be 40.9 and 19.0 corresponding to linear and quadratic fit, respectively.
In addition, to examine whether the quadratic ephemeris is obviously better than the linear ephemeris, we used an analysis of variance (i.e. F-test) presented by Pringle (1975).
The statistic system parameters for linear fit and quadratic fit were computed to be $\lambda_{1}=17.6$ and $\lambda_{2}=55.4$, revealing that the quadratic fit has higher significant level, far in excess of 99.99\%.
Therefore, a quadratic ephemeris is the best description of the general trend of the $O-C$ diagram. In Figure 2, the downward parabola corresponds to a secular decrease in the orbital period at a rate of $\dot{P}={-1.22}\times10^{-11}{days/cycle}=-3.35\times10^{-11}ss^{-1}$.

However, the residuals of the quadratic fit, displayed in Figure 3 (top panel), show an obvious cyclic oscillation. In general, this oscillation arises from the LTT effect via the presence of an unseen circumbinary object. To represent the periodic variation, a common scenario with an eccentric orbit was first considered (e.g., Irwin 1952; Li \& Qian 2014; Qian et al. 2013). However, the eccentricity was determined to be close to zero but with a larger error indicating that the orbit is circular. Therefore, final solutions were obtained by assuming a circular orbit, as follows
\begin{equation}
(O-C)_2=\Delta{T_1}+K\sin(2\pi/P_3\times E+\varphi).
\end{equation}
These parameter values and their explanations were also summarized in Table 2. The $\chi^{2}$ value of the sinusoidal fit is calculated to be $\sim0.7$, indicating a very good fit. Also note that in this case we excluded the timings with the errors larger than 0.0004 days because these errors have been more than the amplitude of periodic oscillation. The result shows that, apart from the long-term decrease, the orbital period of DE CVn also has a cyclic wiggle with an amplitude of 28.08($\pm$5.01) s and a period of 11.22($\pm$0.36) years. In Figure 3, the grey solid line in the upper panel denotes the best-fitting model of periodic oscillation. The lower panel plots the residuals from such sine fitting.

\begin{table}[!h]
\caption{Orbital parameters of the circumbinary companion in DE CVn.}\label{elements}
\begin{center}
\small
\begin{tabular}{llllllllll}
\hline\hline
Parameters                                            &Quadratic ephemeris:  \\
                                                      &$(O-C)_1=\Delta{T_{0}}+\Delta{P_{0}}{E}+{\beta}{E^{2}}$\\
\hline
Revised epoch, ${\Delta{T_{0}}}$ (days)           & $-7.18(\pm3.35)\times10^{-5}$          \\
Revised period, ${\Delta{P_{0}}}$ (days)          & $+7.88(\pm1.50)\times10^{-8}$        \\
Rate of the linear decrease, $2\beta$ (days/cycle)  & $-1.22(\pm0.13)\times10^{-11}$             \\
\hline\hline
Parameters                                            &Sine fitting:  \\
                                                      &$(O-C)_2=\Delta{T_1}+K\sin(2\pi/P_3+\varphi)$\\
\hline
Revised epoch, ${\Delta{T_{1}}}$ (days)           & $9.67(\pm0.32)\times10^{-5}$          \\
The semi-amplitude, $K$ (days)                             & $0.000325(\pm0.000058)$  \\
Orbital period, $P_3$ (years)                                  & $11.22(\pm0.36)$                      \\
The orbital phase, $\varphi$ (deg)                              & $135.55(\pm12.5)$   \\
Projected semi-major axis, $a_{12}\sin{i^{'}}$ $(AU)$            & $0.056(\pm0.010)$  \\
Mass function,$f(m)$ $(M_{\odot})$                                  & $1.42(\pm0.76)\times10^{-6}$         \\
Mass of the third body, $M_{3}\sin{i'}$ $(M_{\odot})$           & $0.011(\pm0.003)$                        \\
Orbital separation, $d_{3}$ ($AU,{i'=90^{\circ}}$)       & $5.75(\pm2.02)$                        \\
\hline\hline
\end{tabular}
\end{center}
\end{table}

\section{Discussions}
\subsection{Physical Causes of Secular Period Variation}

DE CVn is a detached close binary with an orbital period of $\sim8.7$ hr, containing a white dwarf and a low-mass star. In general, the long-term evolution of close, evolved binaries (e.g., PCEBs and CVs) is driven by AMLs. For short-period systems ($\leq2$ hr), the dominant AML mechanism is the emission of GR (e.g., Paczy{\'n}ski 1967; Faulkner 1971; Landau \& Lifshitz 1975), whereas at the longer orbital periods ($\geq3$ hr) the magnetized stellar wind can take away the binary's orbital angular momentum, the so-called MB (e.g., Verbunt \& Zwaan 1981). These processes will cause the binary's orbit to shrink over time. Thus the continuous period decrease of DE CVn may be the result of AML due to GR or/and MB. Note that the GR mechanism is at work in all close binaries, the GR-driven period decrease rate can be calculated by (Kraft et al. 1962; Paczy{\'n}ski 1967)
\begin{equation}
\frac{\dot{P}_{GR}}{P_{orb}}=-3\frac{32G^{3}}{5c^{5}}\frac{M_{1}M_{2}(M_{1}+M_{2})}{a^{4}},
\end{equation}
where $a$ and $P_{orb}$ are the orbital separation and period, respectively. $M_1$ is the primary star's mass and $M_2$ is the secondary star's mass.
The system parameters ($M_1=0.51M_{\odot}$, $M_2=0.41M_{\odot}$) given by van den Besselaar et al. (2007), coupled with Kepler¡¯s third law, derived the separation between two components in binary as $a=2.09R_{\odot}$. In the end, the period decrease rate driven by GR is computed as $\dot{P}_{GR}=-2.60\times{10^{-14}}\,s s^{-1}$,
which is three orders of magnitude smaller than observed one.
To explain the secular change, therefore, we need other AML mechanism, which is most typically seen as MB.

It is generally agreed that the AML rates via MB in the systems with $P_{orb}>3$ hr are well above GR. To estimate the orbital period decay due to MB, we use the standard MB model proposed by Rappaport et al. (1983)
\begin{equation}
\dot{P}_{MB}=-1.4\times10^{-12}(\frac{M_{\odot}}{M_1})(\frac{M_1+M_2}{M_{\odot}})^{1/3}(\frac{R_2}{R_{\odot}})^{\gamma}(\frac{d}{P_{orb}})^{7/3}\, s \, s^{-1},
\end{equation}
where $R_2$ is the secondary star's radius, and $\gamma$ is the magnetic braking index in a range from 0 to 4. The radius of the secondary in DE CVn is $R_2=0.37R_{\odot}$ (van den Besselaar et al. 2007), which can be combined with the standard value of $\gamma=4$ to yield $\dot{P}_{MB}=-5.28\times10^{-13}\,s s^{-1}$. It is about two orders of magnitude smaller than the change seen in Figure 2.
To examine whether the MB mechanism could be responsible for the period decay, we used $\gamma=0$ to maximize the period decrease rate. The maximum period variation via MB is $\dot{P}_{MB}=-2.82\times10^{-11}\,s s^{-1}$, which is also clearly insufficient to explain the true change. Therefore, there should be a more efficient AML mechanism causing the period decay.

Recently, an alternative mechanism for orbital AML in detached binaries, the circumbinary disk model, was proposed (Chen \& Podsiadlowski 2017).
The circumbinary disks may originate from the mass loss during mass transfer (van den Heuvel \& Loore 1973; van den Heuvel 1994),
or are the remaining CE material lost by the white dwarf (Spruit \& Taan 2001). If the circumbinary disk around the binaries, the orbital angular momentum of the system could be efficiently removed by the tidal torques from the disk.
In fact, many studies have shown that the presence of circumbinary disk has a major influence on the binary's evolution such as PCEBs (Chen 2009; Chen \& Podsiadlowski 2017), CVs (Spruit \& Taan 2001; Taan \& Spruit 2001), Algol binaries (Chen et al. 2006) and black-hole X-ray binaries (Chen \& Li 2006, 2015).
Assuming that the detached binary DE CVn is surrounded by a circumbinary disk, the predicted period decrease via the circumbinary disk is given by (Chen \& Podsiadlowski 2017)
\begin{equation}
\dot{P}_{CB}=-6\pi\frac{M_{d}\alpha}{R}(\frac{H}{R})^{2}\frac{a}{\mu},
\end{equation}
where $M_d$ is the circumbinary disk mass, $\alpha$ is the viscosity coefficient, $R$ is the half angular momentum radius of the disk, $H$ is the disk's thickness and $\mu=\frac{M_1M_2}{M_1+M_2}$ is the reduced mass of the system. This circumbinary disk scenario shows that the period change is closely connected to the disk parameters $(M_d\alpha H^2)/R^{3}$ and the properties of the system $a/\mu$.  By using the disk parameters of a PCEB NN Ser ($M_d=2.4\times10^{-4}M_{\odot}$, $\alpha H^2/R^{3}=1.9\times10^{-21}cm^{-1}$) derived by Chen \& Podsiadlowski (2017), they successfully explained the rapid period decrease in seven detached binaries by using this model, and suggested that the circumbinary disk plays a significant role in the PCEB's evolution.
For DE CVn we derive $a/\mu=6.39\times10^{11}cm \, M_{\odot}^{-1}$. This implies that $\dot{P}_{CB}$ in this binary is only determined by the disk parameters.
To test whether the observed period decrease originates from a circumbinary disk, we give in the Figure 4 the $\dot{P}_{CB}$ values for varying disk masses and $\alpha H^2/R^{3}$ according to Equation (6).
The results presented in Figure 4 show that this model is able to interpret the secular period variation of DE CVn safely. Based on the observations and calculations above, the mass of circumbinary disk in DE CVn could be limited in the range of a few$\times$$10^{-4}$-$10^{-3}$$M_{\odot}$,
which is compatible with the derived disk mass from Chen \& Podsiadlowski (2017) in the order of magnitude, and in agreement with the disk mass range observed by Gielen et al. (2007).

Of course, another mechanism cannot entirely be excluded to interpret the apparent period decrease. Alternatively explanation is that the quadratic variation observed in Figure 2 could be only part of a longer-term sinusoidal change caused by a more distant additional companion.

\begin{figure}[!h]
\begin{center}
\includegraphics[width=15cm]{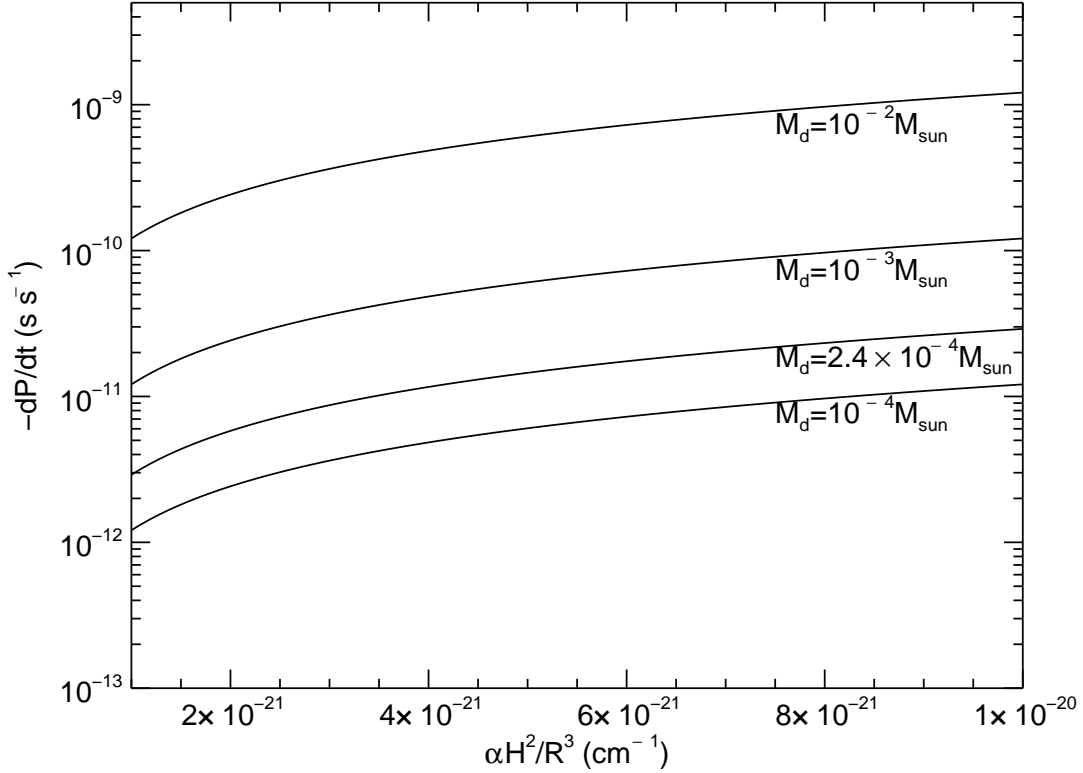}
\caption{Relationship between the theoretical decrease rate in the circumbinary disk model and different disk masses in the $\dot{P}-\alpha H^2/R^{3}$ plot. Typical disk mass for NN Ser ($M_d=2.4\times10^{-4}M_{\odot}$) was also marked in the figure. }
\end{center}
\end{figure}

\subsection{Cyclic Period Variation and Its Possible Interpretations}

The cyclic period changes in PCEBs can be driven by Applegate's mechanism (Applegate 1992) or perturbed by a unseen companion. To examine whether the Applegate mechanism is responsible for the cyclic oscillation, we calculate the required energies to produce this change by using the method of Brinkworth et al. (2006). The analyzing result shows that the donor doesn't have enough energy budget to drive the Applegate process (see Figure 5). Using $T_2=3400$ K for the secondary star with a spectral type of M3V, its luminosity can be computed by $L_{2}=(\frac{R_{2}}{R_{\odot}})^{2}(\frac{T_{2}}{T_{\odot}})^{4}L_{\odot}$. Moreover, the Applegate mechanism in PCEBs has been systematically assessed by V\"{o}lschow et al. (2016), and a modified model also has been employed. These authors suggest that a perfect Applegate PCEB should have a quite close orbit of $\sim0.5R_{\odot}$ with a donor of $\sim0.5M_{\odot}$. This mechanism becomes more reliable for a massive secondary star and a more tight orbit than are presented here. These indicate that the Applegate's mechanism
is not work here and it seems certain that there is a third body orbiting the binary DE CVn.
Our result, however, does not prove the absence of magnetic activity in this system. Instead, it just means that the Applegate's model was not the most important mechanism in this case, the reason is that it cannot contribute significantly to such period wiggle.

The mass of circumbinary companion were derived as $M_{3}\sin{i'}=0.011(\pm0.003)M_{\odot}$, based on the best-fitting parameters and the mass of two components of the binary.
If the orbital inclination of third body is a random distribution, then when $i'\geq51.8^{\circ}$, its mass
is $0.011M_{\odot}\leq M_3\leq0.014M_{\odot}$, probably a giant planet. If $i'\geq8.4^{\circ}$, the mass of third body corresponds to $M_3\leq0.075M_{\odot}$, indicating that it may be a brown dwarf or a planet. Given that the most likely scenario is that both the third body and the eclipsing host star lie in the same plane(i.e., $i^{'}=i=86^{\circ}$),
the third body's mass was estimated to be $M_{3}=0.011M_{\odot}$, and it should be a giant planet.
The orbital separation between this hypothetical giant planet and its host star is about $5.75(\pm2.02)$ AU.

So far, the circumbinary companions have been detected in some PCEBs by using the eclipse timing method. However, the origin of these objects
is quite complex and remains little understood. It is possible that they are first generation
planets formed from a protoplanetary disc material before the CE event, or are second
generation substellar objects formed from the material lost by the white dwarf (V\"{o}lschow et al. 2014; Bear \& Soker 2014; Schleicher \& Dreizler 2014). If they are first generation
companions, how did they survive the CE phase? A study by Mustill et al. (2013) found that they would be difficult to survive during the CE phase due to the orbit stability. For the second-generation scenario, there are still a few problems such as planet-making efficiency (Bear \& Soker 2014). If the planet orbiting DE CVn originated from a second-generation scenario, it remains possible that this binary hold protoplanetary disk material. As mentioned above, the circumbinary disk is most likely formed in the CE material which was not entirely ejected from the system. Therefore, the detection of a circumbinary disk in Section 4.1 may provide some support to the possibility of a second-generation giant planet around DE CVn.
The detection and research of the circumbinary companions and disks in PCEBs may offer some insight into the planetary formation and could enrich our understanding and knowledge on the CE evolution.

\begin{figure}[!h]
\begin{center}
\includegraphics[width=16cm]{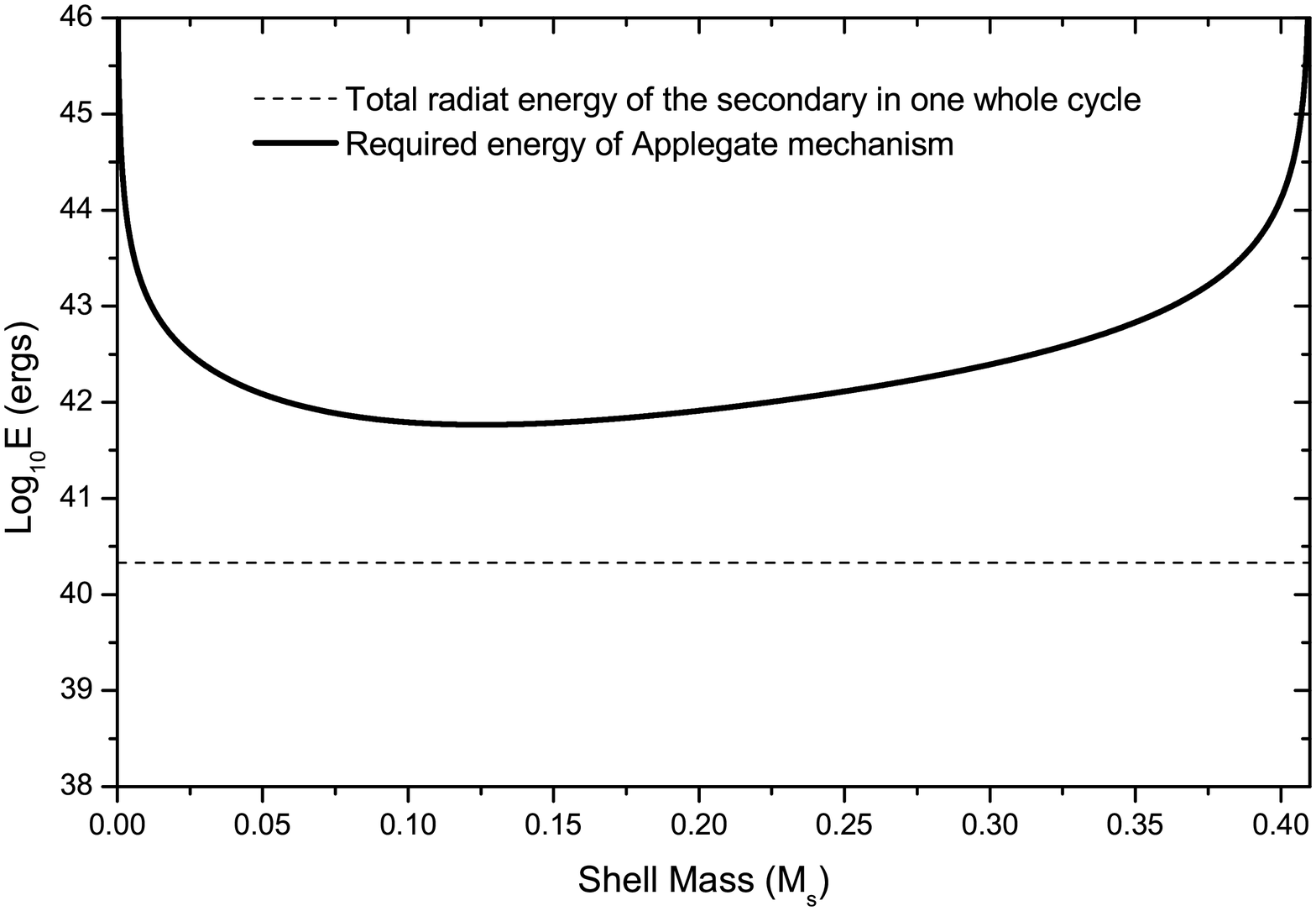}
\caption{The solid line corresponds to the energy needed to generate the discovered period wiggle in DE CVn applying the Applegate's mechanism. $M_s$ represents the supposed shell
mass of the donor. The dashed horizontal line shows all radiant energy from the cool component in a full cycle of the O-C oscillation. }
\end{center}
\end{figure}

\section{Conclusions}

The present work observed and studied the orbital period variations of DE CVn.
We discovered that, besides a long-term decrease, its orbital period also shows a cyclic oscillation.
Generally, the secular period decrease in PCEBs results from the binary's AMLs by GR and MB.
However, the analysis results show that both GR and MB are insufficient to explain the observed decrease rate.
Additional AMLs are required to solve this problem. Based on the investigation of rapid orbital decay in detached binaries, Chen \& Podsiadlowski (2017) indicated that the orbital angular momentum of these systems can be efficiently  extracted by the circumbinary disk. Our investigation believes that the possible mechanism driving the extra AML is
the tidal torque generated from the interaction of the binary and the circumbinary disk. Detailed calculations show that the circumbinary disk model from Chen \& Podsiadlowski (2017) can explain the observed period decrease, indicating that the circumbinary disk play a major role for the evolution of DE CVn. The derived disk mass has the range of a few$\times$$10^{-4}$-$10^{-3}$$M_{\odot}$, which is compatible with the previous conclusions from Gielen et al. (2007) and Chen \& Podsiadlowski (2017). Therefore, the observed period decrease may offer some support for
the presence of circumbinary disk in DE CVn.

Since the secondary star in DE CVn is so feeble to supply the energy needed for the period change,
the cyclic period wiggle can reasonably be interpreted as the gravitational perturbation by a circumbinary object (i.e. the LTT effect).
The mass of the third body was calculated as $M_{3}\sin{i'}=0.011(\pm0.003)M_{\odot}$. Assuming the unseen companion and the eclipsing pair are in the same orbital plane ($i^{'}=i=86^{\circ}$), the mass would match to a giant planet. Recent investigations believe that the PCEBs are one of quite important host
star of the brown dwarfs and planets (e.g., Qian et al. 2015, 2016; Han et al. 2017a). Although some mechanisms have been used to describe the formation of substellar objects, many questions regarding the circumbinary objects in PCEBs still remain.

The results suggest that a circumbinary disk is taking angular momentum from the eclipsing PCEB DE CVn with a giant planet. These make DE CVn a very interesting triple system for further investigating in the future.
As yet, however, the data coverage in this paper
is just 1.5 cycles. That leaves a serious question whether the discovered period wiggle is truly periodic, or merely quasi-periodic.
Also, any proposed circumbinary companions in these systems should be confirmed by other methods such as planetary transits and radial velocity variations.
As for the circumbinary disk, a direct detection in the L-band (3$-$4$\mu m$) is considered to be feasible
because the dust could contribute the continuous spectrum (Spruit \& Taam 2001).
Hence, further multi-waveband observations and investigations of this system are crucially important to confirm our conclusions. In addition, the long-term dynamical stability of this proposed circumbinary disk plus planet system also need to be assessed in future work.

\acknowledgments{This work is partly supported by the National Natural Science Foundation of China (Nos. 11573063, 11611530685, 11803083, U1731238, U1831120), the Key Science Foundation of Yunnan Province (No. 2017FA001) and the Open Project Program of Guizhou Provincial Key Laboratory of Radio Astronomy and Data Processing (No. KF201807). New data presented here were observed with the 60cm, 1m and 2.4m telescopes at the Yunnan Observatories, the 85cm and 2.16m telescopes in Xinglong Observation base in China and the 2.4m Thai National Telescope (TNT) of National Astronomical Research Institute of Thailand (NARIT). We acknowledge the support of the staff of the Xinglong 2.16m and 85cm telescopes. This work was also partially supported by the  Open  Project  Program  of  the  Key  Laboratory  of  Optical  Astronomy,  National Astronomical Observatories, Chinese Academy of Sciences. We appreciate the observers from around the world for the contributions of the AAVSO data of DE CVn. Finally, we would like to thank the anonymous reviewer for the valuable comments and suggestions.}

\end{document}